\documentclass[12pt,preprint]{emulateapj}

\shorttitle{ALMA 0.1--0.2 arcsec resolution imaging of NGC 1068}   
\shortauthors{Imanishi et al.}

\begin{document}


\title{ALMA 0.1--0.2 arcsec resolution imaging of the NGC 1068 nucleus 
- compact dense molecular gas emission at the putative AGN location}


\author{Masatoshi Imanishi \altaffilmark{1,2}}
\affil{Subaru Telescope, 650 North A'ohoku Place, Hilo, Hawaii, 96720,
U.S.A.} 
\email{masa.imanishi@nao.ac.jp}

\author{Kouichiro Nakanishi \altaffilmark{2}}
\affil{National Astronomical Observatory of Japan, 2-21-1
Osawa, Mitaka, Tokyo 181-8588, Japan}

\and 

\author{Takuma Izumi}
\affil{Institute of Astronomy, School of Science, The University of Tokyo,
2-21-1 Osawa, Mitaka, Tokyo 181-0015, Japan}

\altaffiltext{1}{National Astronomical Observatory of Japan, 2-21-1
Osawa, Mitaka, Tokyo 181-8588, Japan}
\altaffiltext{2}{Department of Astronomical Science,
The Graduate University for Advanced Studies (SOKENDAI), 
Mitaka, Tokyo 181-8588, Japan}

\begin{abstract}
We present the results of our ALMA Cycle 2 high angular resolution
(0$\farcs$1--0$\farcs$2) observations of the nuclear region of the nearby 
well-studied type-2 active galactic nucleus (AGN), NGC 1068, at HCN
J=3--2 and HCO$^{+}$ J=3--2 emission lines. 
For the first time, due to a higher angular resolution than previous studies, 
we clearly detected dense molecular gas emission
at the putative AGN location, identified as a $\sim$1.1 mm ($\sim$266
GHz) continuum emission peak, by separating this emission from brighter
emission located at  
0$\farcs$5--2$\farcs$0 on the eastern and western sides of the AGN. 
The estimated intrinsic molecular emission size and dense molecular mass,
which are thought to be associated with the putative dusty molecular
torus around an AGN, were $\sim$10 pc and $\sim$several $\times$
10$^{5}$M$_{\odot}$, respectively.
HCN-to-HCO$^{+}$ J=3--2 flux ratios substantially higher than unity
were found throughout the nuclear region of NGC 1068. 
The continuum emission displayed an elongated morphology along the
direction of the radio jet 
located at the northern side of the AGN, as well as a weak 
spatially resolved component at $\sim$2$\farcs$0 on the southwestern
side of the AGN. 
The latter component most likely originated from star formation,
with the estimated luminosity more than one order of magnitude lower than
the luminosity of the central AGN.  
No vibrationally excited (v$_{2}$=1f) J=3--2 emission lines were detected
for HCN and HCO$^{+}$ across the field of view.
\end{abstract}


\section{Introduction}

An active galactic nucleus (AGN) produces very bright emission from the
central compact region of its galaxy due to the release of energy by a
mass-accreting supermassive black hole. 
AGNs can be classified as either type-1 (which show broad optical
emission lines) or type-2 (which do not). 
Based on observations of the nearby type-2 AGN, NGC 1068 (z = 0.0037,
distance = 14 Mpc, 1 arcsec is $\sim$70 pc), the widely accepted unified
picture of AGNs was proposed, which postulates that a central engine 
(i.e. a mass-accreting supermassive black hole) is surrounded by a
toroidally distributed dusty medium, the so-called ``dusty torus''
\citep{ant85}.  
Dust located in close proximity to a central AGN is heated by the AGN's
energetic radiation and produces strong
mid-infrared (3--20 $\mu$m) emission. 
Thus, high angular resolution mid-infrared observations of AGNs are a 
powerful tool for understanding the properties of the dusty torus.

Based on nuclear infrared to submillimeter photometric observations and
a clumpy torus model, the outer size of the dusty torus of NGC 1068 is
estimated to be 20$^{+6}_{-10}$ pc \citep{gar14}.
Very high angular resolution ($\sim$10 mas) mid-infrared 
($\sim$10 $\mu$m) interferometric imaging is another powerful tool that can
directly constrain the size of the mid-infrared-emitting region in the
dusty torus and was applied to NGC 1068 to derive its size of
$<$several pc \citep{jaf04,pon06,rab09,bur13,lop14}.  
Because mid-infrared $\sim$10 $\mu$m observations selectively trace hot
($>$a few 100 K) dust in the inner part of the torus and are not
sensitive to cooler dust in the outer part, the actual size of the dust
torus could be larger than estimated.  
Where dust is present in the universe, molecular gas usually
co-exists. 
As the density of dust is likely to be enhanced in the dusty torus,
compared with the surrounding area, dense gas 
($>$10$^{4}$--10$^{5}$ cm$^{-3}$) is expected to be abundant  
in this region.  
High angular resolution molecular gas observations of the nucleus of NGC
1068, using (sub)millimeter interferometry and dense gas tracers 
such as HCN and HCO$^{+}$, can play an important role in observationally
constraining the properties of the dusty molecular torus by
reducing contamination from more diffuse, less dense molecular gas
emission from the host galaxy.  

Using the Submillimeter Array (SMA) with
an angular resolution of 0$\farcs$5--2$\farcs$0, HCN J=3--2, HCO$^{+}$
J=3--2 and J=4--3 emission lines were detected at the nucleus of NGC 1068 
\citep{kri11}. 
Subsequent Atacama Large Millimeter/submillimeter Array (ALMA) higher 
angular resolution (0$\farcs$5--0$\farcs$6) 
observations further scrutinized the morphology of the molecular gas
emission and revealed that the emission primarily originates from the
eastern and western knots, 
0$\farcs$5--2$\farcs$0 arcsec (35--70 pc) offset from the putative AGN
position \citep{gar14}. 
The properties of compact molecular gas emission associated with the
dusty molecular torus are not yet clear observationally, because the
emission is overwhelmed by the much brighter nearby eastern knot
\citep{gar14}. Molecular gas observation with even higher angular
resolution may better reveal the molecular emission arising from the dusty
molecular torus.
In this paper, we present the results of our ALMA Cycle 2 observations
of the nucleus of NGC 1068 at HCN J=3--2 and HCO$^{+}$ J=3--2 lines,
with an angular resolution of 0$\farcs$1$\times$0$\farcs$2. 

\section{Observations and data analysis}

Band 6 (211--275 GHz) ALMA observations of the nucleus of NGC 1068 
were conducted as part of our Cycle 2 program 2013.1.00188.S (PI =
M. Imanishi) on 2015 September 19 (UT), with a median precipitable 
water vapor value of $\sim$1.9 mm.
In our proposal, we had requested an angular resolution 
of $<$0$\farcs$6,
and we adopted the widest 1.875 GHz width mode in each spectral window.
There were 36 antennas, and the baseline lengths were 41--2270 m.
J0224$+$0659, J0334$-$401, and J0219$+$0120 were used as bandpass, flux,
and phase calibrators, respectively. 
The total net on-source integration time was 24.9 min.    

We covered 263.9--268.6 GHz with three spectral windows, allowing us to
observe HCN J=3--2 ($\nu_{\rm rest}$ = 265.886 GHz) in one spectral
window and HCO$^{+}$ J=3--2 ($\nu_{\rm rest}$ = 267.558 GHz) in another
window.  
HCN v$_{2}$=1f J=3--2 ($\nu_{\rm rest}$ = 267.199 GHz) was observed in
the same spectral window as HCO$^{+}$ J=3--2. 
The HCO$^{+}$ v$_{2}$=1f J=3--2 ($\nu_{\rm rest}$ = 268.689 GHz) line was
covered in the third spectral window.  

Data analysis began with calibrated data provided by the Joint
ALMA Observatory, using CASA 4.5.0 (https://casa.nrao.edu).
After checking the visibility plots, we estimated the continuum flux
level using channels that were not affected by strong emission
lines.     
We subtracted the estimated constant continuum level using the 
CASA task ``uvcontsub'', and then applied the task ``clean'' 
(natural weighting, gain $=$ 0.1, threshold $=$ 0.8 mJy) to create
final continuum-subtracted molecular emission line data.
The clean task  was also applied to the extracted continuum data 
(the so-called ``mfs'' (Multi-frequency synthesis)  method).  
During the clean procedures, we binned 40 channels 
($\sim$20 km s$^{-1}$) to
increase the signal-to-noise ratios and utilized a pixel scale 
of 0$\farcs$03
pixel$^{-1}$, because our observations were conducted with a long
maximum baseline. 
As we intended to discuss emission morphology not only at the 
nucleus, but also at some off-nuclear regions, 
we applied the primary beam correction to all data.
The absolute flux calibration uncertainty and maximum recoverable scale 
of our ALMA band 6 data are $<$30\% 
\footnote{
This is mainly caused by the flux uncertainty of the flux calibrator,
J0334$-$401.
}
and $\sim$4$\farcs$0, 
respectively.  

Because we made only one measurement of the NGC 1068 nucleus, and the 
field of view of our ALMA band 6 data is $\sim$10 arcsec in radius, dense
molecular gas emission from the well-known star-forming molecular gas
ring at 10--15 arcsec from the nucleus \citep{tsa12,gar14} was not
adequately probed.   
Thus, our discussion focuses on the nuclear dense molecular gas emission
properties within a central region of a few arcsec. 
We have confirmed that the PSFs of our data are fairly clean, with
no significant side lobes are recognizable at the region of our interest.

\section{Results}

Figure 1 shows a map of the continuum. 
The achieved angular resolution was 0$\farcs$23$\times$0$\farcs$12,
which was significantly higher than our specified minimum.
The peak flux is estimated to be 9.2 mJy beam$^{-1}$ (51$\sigma$) at 
(02$^{h}$42$^{m}$40$_{.}$711$^{s}$, $-$00$^{\circ}$00$'$47.93$"$)J2000
in our 0$\farcs$03 pixel$^{-1}$ continuum map. The peak coordinate
agrees with the reported AGN location 
at (02$^{h}$42$^{m}$40.71$^{s}$, $-$00$^{\circ}$00$'$47.94$''$)J2000 
\citep{roy98,gar14}.  
We therefore regard the coordinates of our continuum peak as the AGN location, 
denoted as ``C-peak''.

Integrated intensity (moment 0) maps for the HCN J=3--2 and HCO$^{+}$ J=3--2
emission lines are presented in Fig. 2, and their properties are
summarized in Table 1. 
The angular resolution in these maps is 
0$\farcs$1--0$\farcs$2.
The brightest HCN J=3--2 and HCO$^{+}$ J=3--2 emission {\bf is} found
at $\sim$1$\farcs$0 on the eastern side of the AGN (C-peak), with a peak
position at 
(02$^{h}$42$^{m}$40.77$^{s}$, $-$00$^{\circ}$00$'$47.87$''$)J2000, 
which is denoted as ``E-peak''.  
Additionally, weak emission is observed at the western side of the
AGN; here, the peak positions are slightly different between HCN and
HCO$^{+}$.
We tentatively denote the HCO$^{+}$ J=3--2 emission peak in the western
knot at (02$^{h}$42$^{m}$40.63$^{s}$, $-$00$^{\circ}$00$'$48.24$''$)J2000 
as ``W-subpeak''.
These properties have already been recognized in previously obtained
interferometric HCN and HCO$^{+}$ maps \citep{kri11,gar14}. 
Our high angular resolution (0$\farcs$1--0$\farcs$2) maps allow
compact molecular emission to be clearly identified at the AGN location
(C-peak) for the first time, by separating this emission from the
brighter molecular gas emission from the eastern and western sides of
the AGN.  

The spectra at E-peak, W-subpeak, and C-peak are shown in Fig. 3.
To investigate the overall molecular line emission properties within
the central region of NGC 1068 of a few arcsec, we created a spectrum by
integrating a 2$\farcs$4 radius circular region around 
(02$^{h}$42$^{m}$40.69$^{s}$, $-$00$^{\circ}$00$'$48.50$''$)J2000, 
as shown in Fig. 3. 

Gaussian fits for the HCN J=3--2 and HCO$^{+}$ J=3--2 emission lines 
in the spectra, within the beam size, at the E-peak, W-subpeak, and
C-peak are also shown in Fig. 2.
Molecular line fluxes, estimated from the Gaussian fits, within the
beam size are summarized in Table 1. 
Gaussian fits to the detected molecular emission lines in the
area-integrated spectrum (2$\farcs$4 radius circular region) and their
best-fit parameters are included in Fig. 2 and Table 1, respectively.
We adopt the fluxes estimated with these Gaussian fits for our subsequent
discussion. 
A broad emission component is detected at the C-peak, but not in the
other regions.  
This broad component may be related to highly turbulent molecular
gas in the torus \citep{wad09} and/or possible compact outflow activity. 

Figure 4 presents intensity-weighted mean velocity (moment 1) maps for the
HCN J=3--2 and HCO$^{+}$ J=3--2 emission lines, to provide a better
understanding of their dynamics. 

In the spectra, we see no clear signatures of vibrationally excited
v$_{2}$=1f J=3--2 emission lines of HCN and HCO$^{+}$. 
We generated integrated-intensity (moment 0) maps of these lines by
summing ten velocity channels with v$_{\rm opt}$ = 1000--1210 
km s$^{-1}$, but obtained
non-detection with 3$\sigma$ upper limits of $<$0.15 
[Jy beam$^{-1}$ km s$^{-1}$], both for vibrationally excited
(v$_{2}$=1f) HCN J=3--2 and HCO$^{+}$ J=3--2 emission lines.  

\section{Discussion}

\subsection{Molecular emission from the putative AGN dusty torus}

Compact HCN J=3--2 and HCO$^{+}$ J=3--2 emission lines are clearly
detected at the C-peak, where the putative dusty molecular torus is
located.
The most natural explanation is that this compact emission is associated 
with the dusty molecular torus around an AGN's central engine.
The deconvolved intrinsic size of the HCN J=3--2 emission, estimated using
the CASA task ``imfit'', is 175$\pm$52 [mas] $\times$ 92$\pm$31
[mas], or 12.3$\pm$3.6 [pc] $\times$ 6.2$\pm$2.2 [pc]  
\footnote{
The size measurement of the HCO$^{+}$ J=3--2 emission at the C-peak, using 
``imfit'', provides 164$\pm$94 [mas] $\times$ 54$\pm$121 [mas].}.
This is roughly comparable to the estimated torus size of 20$^{+6}_{-10}$
[pc], based on a nuclear infrared to submillimeter spectral energy
distribution and a clumpy torus model \citep{gar14}. 

We estimated the mass of dense molecular gas in the dusty molecular
torus from the detected compact HCN J=3--2 emission at the C-peak,
within the beam size.
If HCN emission is thermalized at up to J=3 and optically thick, then the HCN
J=1--0 flux is 1/9 of the HCN J=3--2 flux. 
If the emission is sub-thermal, then the HCN J=1--0 flux is larger than 1/9 
of the HCN J=3--2 flux.
The HCN J=1--0 luminosity, derived using Equations (1)
and (3) of \citet{sol05}, is 2.0$\pm$0.6 [L$_{\odot}$] or
(8.8$\pm$2.6)$\times$10$^{4}$ [K km s$^{-1}$ pc$^{2}$], or possibly even
higher.
Adopting the relationship between dense molecular mass and HCN J=1--0
luminosity, M$_{\rm dense}$ = 10 $\times$ HCN J=1--0 luminosity 
[M$_{\odot}$ (K km s$^{-1}$ pc$^{2}$)$^{-1}$] \citep{gao04,kri08},
we obtain a dense molecular mass in the torus of several 
$\times$ 10$^{5}$ M$_{\odot}$ (with large potential uncertainty), 
which roughly agrees to the gas mass estimated with the clumpy torus model, 
(2.1$\pm$1.2)$\times$10$^{5}$ M$_{\odot}$ \citep{gar14}. 
This agreement suggests that the bulk of the molecular gas in the torus is
in a dense form and has been properly probed with our HCN J=3--2 observation.

In the HCN/HCO$^{+}$ J=3--2 moment 0 map, faint emission
connecting the C-peak to the bright eastern molecular knot is 
just visible in the northeastern direction from the C-peak. 
The northern molecular gas is thought to be further out than the central AGN
and its putative dusty torus \citep{gar14}.
The connected part is the least redshifted in the northern region 
(i.e., the observed velocity toward the other side is the smallest) and may
be related to the molecular gas supply from the eastern molecular region
to the central AGN torus, though higher sensitivity observations are
necessary to pursue this scenario in further detail.  

\subsection{Continuum emission}

The continuum emission in Fig. 1 displays an elongated morphology along
the northern direction from the peak. 
As the putative dusty molecular torus is estimated to be oriented along the 
southeastern to northwestern direction \citep{gre96,rab09}, 
the continuum elongation is visible in a direction that is not
intercepted by the torus.  
Synchrotron emission along the radio jet \citep{roy98,gal04} and thermal
radiation from mid-infrared 10--20 $\mu$m continuum-emitting dust
\citep{alo00,tom01} also show a similarly elongated morphology along the
northern direction. 
Multi-frequency 0$\farcs$1--0$\farcs$2 angular resolution ALMA data will
help to resolve the origin of this elongated continuum emission.

Spatially extended weak emission is visible at $\sim$2$\farcs$0 in the
southwestern part of the C-peak or the southern part of the W-subpeak, 
at  
(02$^{h}$42$^{m}$40.60--40.65$^{s}$, $-$00$^{\circ}$00$'$48.5--49.7$''$)
J2000. 
In this region, similarly spatially extended HCN J=3--2 and HCO$^{+}$
J=3--2 emission lines are also evident in Fig. 2.
Continuum emission at 349 GHz and 689 GHz was also detected in this area
\citep{gar14}.
A natural origin for the continuum emission at $\sim$260 GHz 
is star-formation, because newly-born stars in dense molecular gas
can (1) heat the surrounding dust and emit dust thermal radiation, and 
(2) create HII-regions, from which thermal free-free emission is
produced.  
In fact, in Fig. 2, HCN J=3--2 and HCO$^{+}$ J=3--2 emission lines are
weaker in the region where continuum emission is detected (hereafter, the SW 
star-forming region), than the region close to the W-subpeak. 
It is likely that conversion from dense molecular gas to stars and 
additional possible stellar feedback to the surrounding molecular gas
dissipate the dense molecular gas and weaken the HCN J=3--2 and
HCO$^{+}$ J=3--2 emission lines in the SW star-forming region.

We measured the total continuum flux in the SW star-forming region
(rectangular with a width of 0$\farcs$7 in the east-west $\times$ a
width of 2$\farcs$1 in the north-south) (02$^{h}$42$^{m}$40.59--40.64$^{s}$, $-$00$^{\circ}$00$'$47.89--50.00$''$)J2000 to be $\sim$10 mJy.  
If this is solely due to the thermal free-free emission inside
star-forming HII-regions, then the corresponding far-infrared (40--500
$\mu$m) luminosity becomes $\sim$6 $\times$ 10$^{43}$ [erg s$^{-1}$],
or there is a star formation rate of $\sim$3 M$_{\odot}$ yr$^{-1}$
\citep{ken98}, when Equation (1) of \citet{nak05} is used.
If dust thermal radiation contributes in some way to the continuum flux
there, the estimated star-formation luminosity will be smaller. 
Previous infrared spectroscopy failed to clearly detect the signatures
of ongoing active star-formation at the central few arcsec region of the NGC
1068 nucleus, due to the lack of polycyclic aromatic hydrocarbon 
(PAH) emission features \citep{ima97,lef01}, which are a good probe of
star-formation activity \citep{moo86,imd00}. 
It was also argued that the molecular gas properties at the NGC 1068 nucleus
are dominated by AGN radiation, rather than star-formation
\citep{use04,gar10}. 
Thus, our continuum emission map provides a new signature for
the detectable amount of star-formation activity in the nuclear few arcsec
region of NGC 1068, thanks to the high sensitivity of ALMA.   
The PAH emission flux within the nuclear 3$\farcs$8 $\times$ 3$\farcs$8
region in infrared spectroscopy is estimated to be $<$2.7 $\times$
10$^{40}$ [erg s$^{-1}$] \citep{ima02}, which corresponds to
star-formation-originated far-infrared luminosity with $<$2.7 $\times$
10$^{43}$ [erg s$^{-1}$] \citep{mou90}.  
This upper limit is apparently smaller than the above estimate but can
be reconciled because (1) not all of the SW star-forming region was covered 
by previous infrared slit spectroscopy \citep{ima97}, (2) the estimated
star-formation luminosity is reduced if dust thermal radiation
contributes to the observed continuum emission at $\sim$266 GHz, and 
(3) some fraction of the PAHs can potentially be destroyed by the AGN's strong
X-ray radiation in the close vicinity of the AGN in NGC 1068
\citep{voi92}.    

\subsection{Molecular emission at the nuclear region}

Throughout the nuclear few arcsec region of NGC 1068, the HCN J=3--2 flux
tends to be higher than the HCO$^{+}$ J=3--2 flux in the moment 0 maps.
We compared the HCN J=3--2 and HCO$^{+}$ J=3--2 fluxes, based on
Gaussian fits, at representative regions, E-peak, W-subpeak, C-peak
(all within the beam size), and the 2$\farcs$4 radius circular region.  
The HCN-to-HCO$^{+}$ J=3--2 flux ratios were 1.6--3.3 (Table 1), and were much
higher than those of the starburst regions \citep{ima16a}.
Enhanced HCN-to-HCO$^{+}$ flux ratios were also found in the nucleus of NGC
1068 at J=1--0 \citep{kri08} and J=4--3 \citep{gar14,vit14}.
Although signatures of the SW star-forming region are found in our ALMA
data, the estimated star-formation luminosity of L$_{\rm SF-SW}$ $<$ 6
$\times$ 10$^{43}$ [erg s$^{-1}$] is an order of magnitude lower than
the central AGN luminosity of L$_{\rm AGN}$ $\sim$ 6 $\times$ 10$^{44}$ 
[erg s$^{-1}$] \citep{boc00}.
It has been argued that AGNs tend to have higher HCN-to-HCO$^{+}$
J-transition flux ratios than starburst regions
\citep{koh05,kri08,ima07,ima09,ima14,izu16}.   
Our observational results of high HCN-to-HCO$^{+}$ J=3--2 flux ratios at
the NGC 1068 nuclear region can naturally be explained by the effect of
the AGN. 


\section{Summary}

We conducted 0$\farcs$1--0$\farcs$2 angular-resolution ALMA Cycle 2
observations of the nucleus of NGC 1068 at HCN J=3--2, HCO$^{+}$ J=3--2,
and nearby continuum emission. 
The high angular resolution of our data allowed detection of dense
molecular gas emission at the putative AGN dusty molecular torus for the
first time, by clearly separating this emission from brighter emission at
0$\farcs$5--2$\farcs$0 on the eastern and western sides of the central AGN.
Enhanced HCN-to-HCO$^{+}$ J=3--2 flux ratios, substantially larger than
unity, were found throughout the region a few arcsec from the nucleus of
NGC 1068. We detected continuum emission, most likely to have originated
during star formation, at $\sim$2$\farcs$0 in the southwestern part of
the AGN.  
These are the first clearly resolved ongoing star-formation signatures,
despite much lower luminosity than the AGN, in the NGC 1068 nuclear region. 

We thank the anonymous referee for his/her useful comment which
helped to improve the clarity of this manuscript.
We are grateful to Dr. Y. Ao and A. Kawamura for their supports
regarding ALMA data analysis. 
M.I. was supported by JSPS KAKENHI Grant Number 23540273 and 15K05030.
This paper makes use of the following ALMA data:
ADS/JAO.ALMA\#2013.1.00188.S. 
T.I. is thankful for the fellowship received from the Japan Society
for the Promotion of Science (JSPS).
ALMA is a partnership of ESO (representing its member states), NSF (USA) 
and NINS (Japan), together with NRC (Canada), NSC and ASIAA
(Taiwan), and KASI (Republic of Korea), in cooperation with the Republic
of Chile. The Joint ALMA Observatory is operated by ESO, AUI/NRAO, and
NAOJ. 

\clearpage

\begin{deluxetable}{lc|clc|llcc}
\tabletypesize{\scriptsize}
\tablecaption{Properties of detected molecular emission lines
\label{tbl-2}}  
\tablewidth{0pt}
\tablehead{
\colhead{Position} & \colhead{Line} & 
\multicolumn{3}{c}{Integrated intensity (moment 0) map} &  
\multicolumn{4}{c}{Gaussian line fit} \\  
\colhead{} & \colhead{[GHz]} &\colhead{Peak} &
\colhead{rms} & \colhead{Beam} & \colhead{Velocity} 
& \colhead{Peak} & \colhead{FWHM} & \colhead{Flux} \\ 
\colhead{} & \colhead{} & 
\multicolumn{2}{c}{[Jy beam$^{-1}$ km s$^{-1}$]} & 
\colhead{[$''$ $\times$ $''$] ($^{\circ}$)} &
\colhead{[km s$^{-1}$]} & \colhead{[mJy]} & \colhead{[km s$^{-1}$]} &
\colhead{[Jy km s$^{-1}$]} \\  
\colhead{(1)} & \colhead{(2)} & \colhead{(3)} & \colhead{(4)} & 
\colhead{(5)} & \colhead{(6)} & \colhead{(7)} & \colhead{(8)} &
\colhead{(9)} 
}
\startdata 
E-peak & HCN J=3--2 & 5.8(18$\sigma$) & 0.32 &
0.22$\times$0.13 (54$^{\circ}$) & 1090$\pm$2 & 28$\pm$1 & 193$\pm$4 & 
5.7$\pm$0.1 \\ 
 & HCO$^{+}$ J=3--2 & 2.6 (17$\sigma$) & 0.15 & 
0.24$\times$0.13 (57$^{\circ}$) & 1097$\pm$3 & 13$\pm$1 & 186$\pm$7 &
2.6$\pm$0.1 \\   
W-subpeak & HCN J=3--2 & 1.4(4.5$\sigma$) & 0.32 & 
0.22$\times$0.13 (54$^{\circ}$) & 1165$\pm$4 & 9.2$\pm$0.4 & 152$\pm$9 &
1.5$\pm$0.1\\     
 & HCO$^{+}$ J=3--2 & 0.65 (4.4$\sigma$) & 0.15 &
0.24$\times$0.13 (57$^{\circ}$) & 1159$\pm$8 & 4.1$\pm$0.4 & 147$\pm$20
& 0.64$\pm$0.11 \\   
C-peak & HCN J=3--2 & 1.1(3.4$\sigma$) & 0.32 &
0.22$\times$0.13 (54$^{\circ}$) & 1110$\pm$7, 1151$\pm$33 & 3.8$\pm$0.7,
1.9$\pm$0.4 & 69$\pm$16, 411$\pm$76 & 1.1$\pm$0.3 \\  
 & HCO$^{+}$ J=3--2 & 0.66(4.5$\sigma$) & 0.15 &
0.24$\times$0.13 (57$^{\circ}$) & 1125$\pm$12, 1081(fix) & 2.9$\pm$0.8,
0.83$\pm$0.38 & 81$\pm$21, 474(fix) & 0.67$\pm$0.22 \\  
2$\farcs$4 & HCN J=3--2 & --- & --- & --- & 
1101$\pm$1 & 640$\pm$8 & 212$\pm$3 & 144$\pm$3 \\     
 & HCO$^{+}$ J=3--2 & --- & --- & --- & 1101$\pm$3 &
243$\pm$7 & 171$\pm$6 & 44$\pm$2 \\ 

\enddata

\tablecomments{ 
Col.(1): Position. 
Col.(2): Line.
Col.(3): Integrated intensity in [Jy beam$^{-1}$ km s$^{-1}$] at the 
emission peak. 
The detection significance relative to the rms noise (1$\sigma$) in the 
moment 0 map is shown in parentheses. 
Possible systematic uncertainty is not included. 
For HCN J=3--2 and HCO$^{+}$ J=3--2, 22 and 21 velocity channels 
of $\sim$20 MHz ($\sim$20 km s$^{-1}$) width for each 
(870--1350 km s$^{-1}$ and 890--1350 km s$^{-1}$) are summed, 
respectively.
Col.(4): rms noise (1$\sigma$) level in the moment 0 map in 
[Jy beam$^{-1}$ km s$^{-1}$], derived from signals in the annular region
with 2$\farcs$4--6$\farcs$0 centered at (02$^{h}$42$^{m}$40.70$^{s}$, 
$-$00$^{\circ}$00$'$48.5$''$)J2000, to avoid an area of significant 
molecular emission line detection.  
Col.(5): Beam size in [arcsec $\times$ arcsec] and position angle in
[degree]. Position angle is 0$^{\circ}$ along the north-south direction, 
and increases counterclockwise. 
Cols.(6)--(9): Gaussian fits of emission lines in the spectra.
At the C-peak, two Gaussian fits are made because a very broad
component is observed, in addition to a narrow component.  
Col.(6): Optical velocity (v$_{\rm opt}$) of emission peak in [km s$^{-1}$]. 
Col.(7): Peak flux in [mJy]. 
Col.(8): Observed FWHM in [km s$^{-1}$] in the spectra of Fig. 2.  
Col.(9): Flux in [Jy km s$^{-1}$].}

\end{deluxetable}

\begin{figure}
\begin{center}
\includegraphics[angle=0,scale=.62]{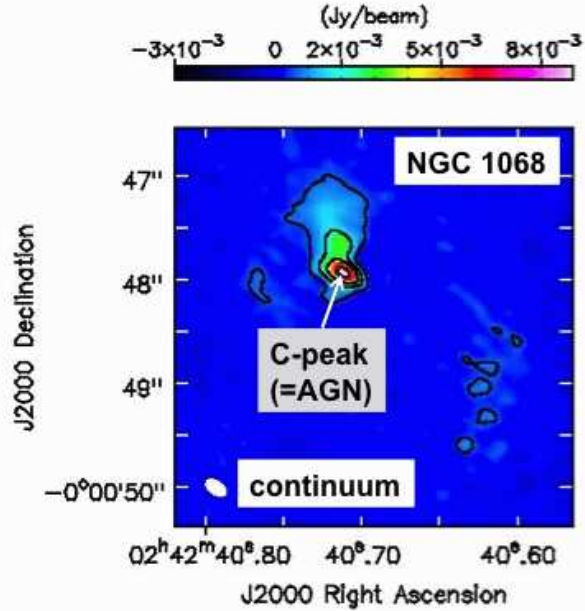} 
\end{center}
\caption{Continuum emission map with rms noise $\sim$ 0.18 mJy beam$^{-1}$. 
The plotted contours are 3$\sigma$, 10$\sigma$, 20$\sigma$, and
40$\sigma$.}
\end{figure}

\clearpage

\begin{figure}
\begin{center}
\includegraphics[angle=0,scale=.56]{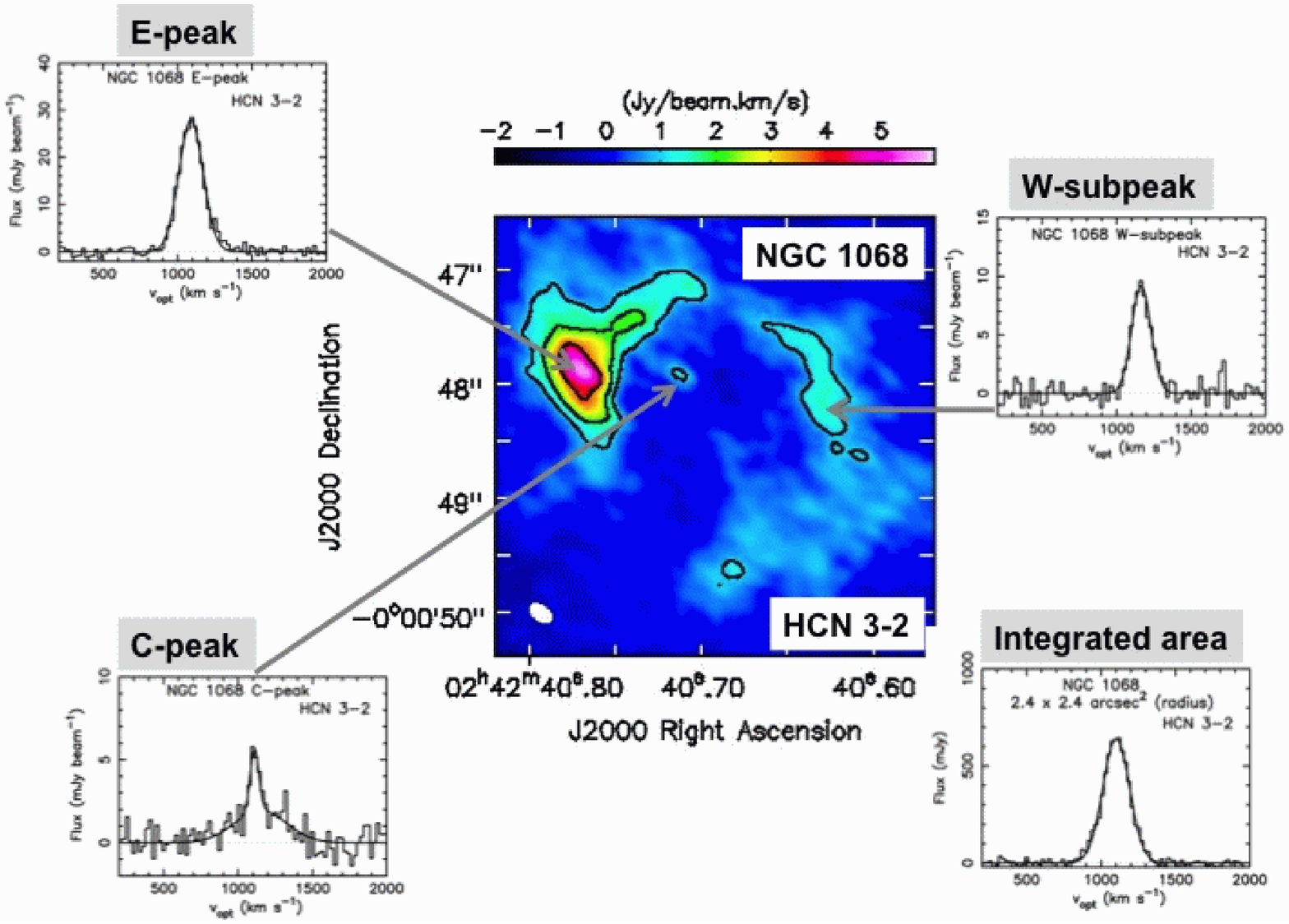} 
\includegraphics[angle=0,scale=.56]{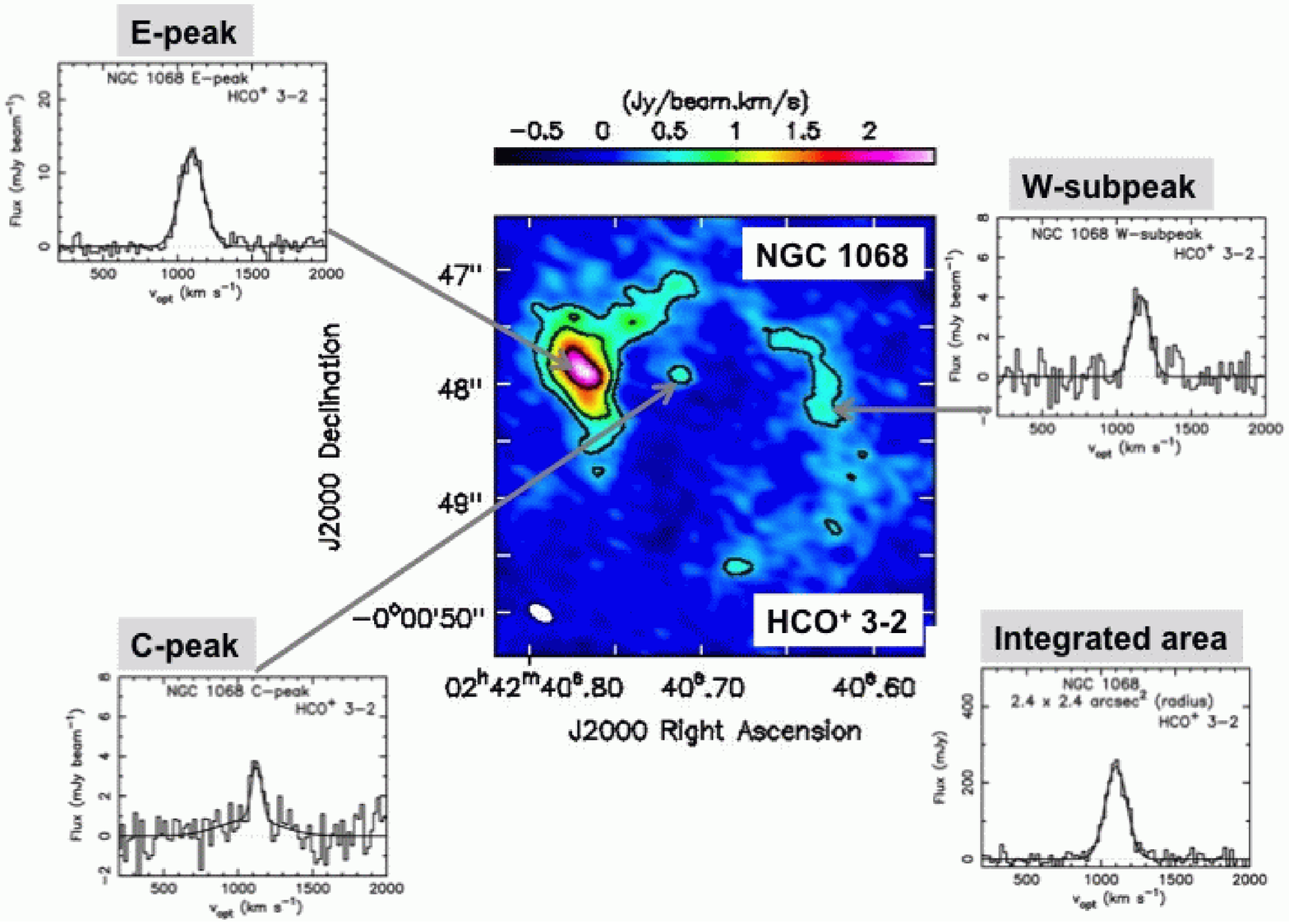} \\
\end{center}
\caption{Integrated intensity (moment 0) maps of HCN J=3--2 and 
HCO$^{+}$ J=3--2 emission lines.
The contours are 3$\sigma$, 6$\sigma$, 12$\sigma$ for HCN J=3--2 and
HCO$^{+}$ J=3--2.
Gaussian fits to molecular emission lines detected in the spectra, 
within the beam size, at individual locations, E-peak, W-subpeak,
C-peak, are also shown. 
The abscissa and ordinate are optical velocity (v$_{\rm opt}$) 
in [km s$^{-1}$] and flux density in [mJy beam$^{-1}$], respectively.
The rms is $\sim$1 mJy beam$^{-1}$, and the Jy-to-K conversion factor is
$\sim$600 [K Jy$^{-1}$] for our observed frequency and beam size.
Gaussian fits in the spectra of an integrated area with a 2$\farcs$4
radius circular region are also added.
}
\end{figure}

\begin{figure}
\begin{center}
\includegraphics[angle=0,scale=.4]{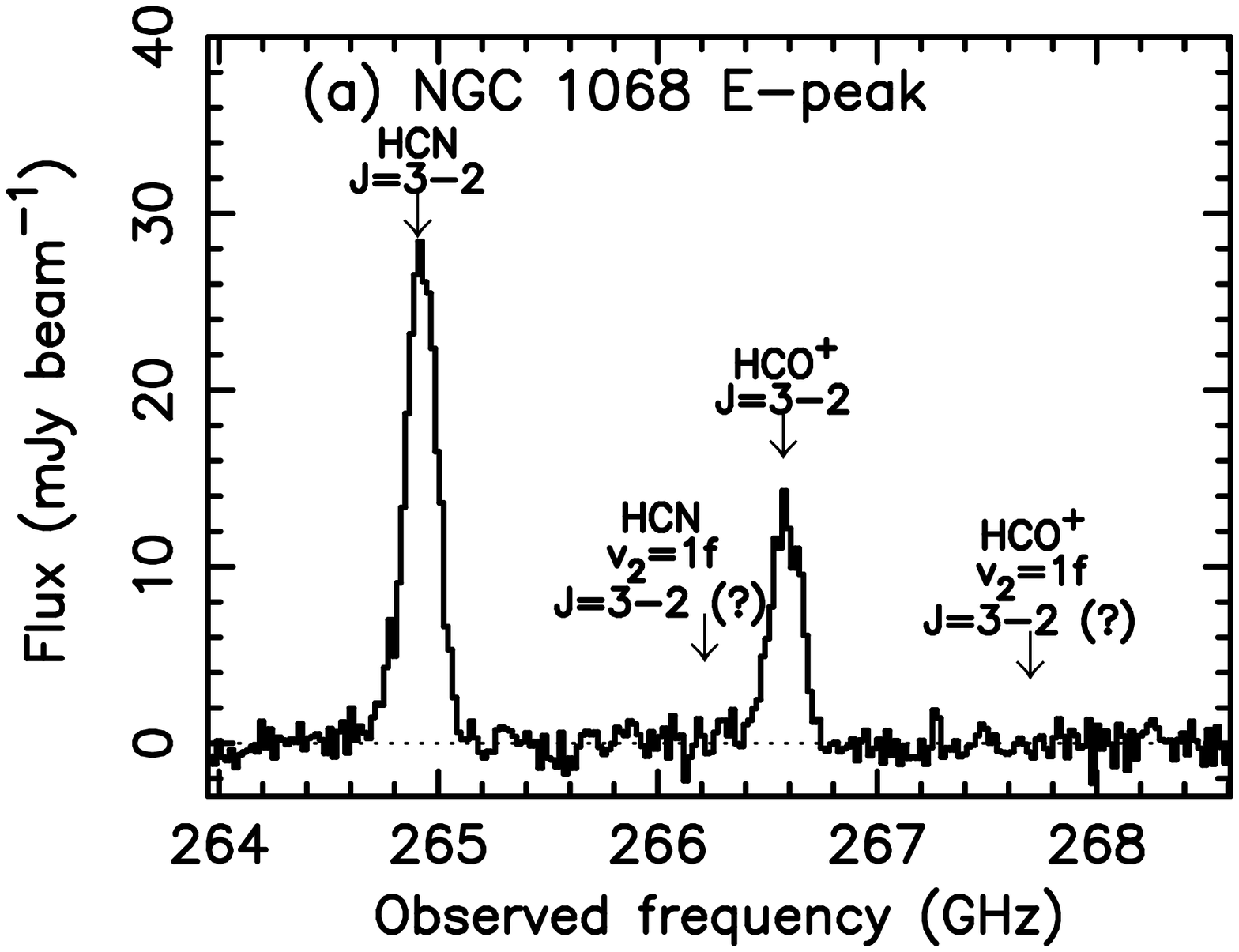} 
\includegraphics[angle=0,scale=.4]{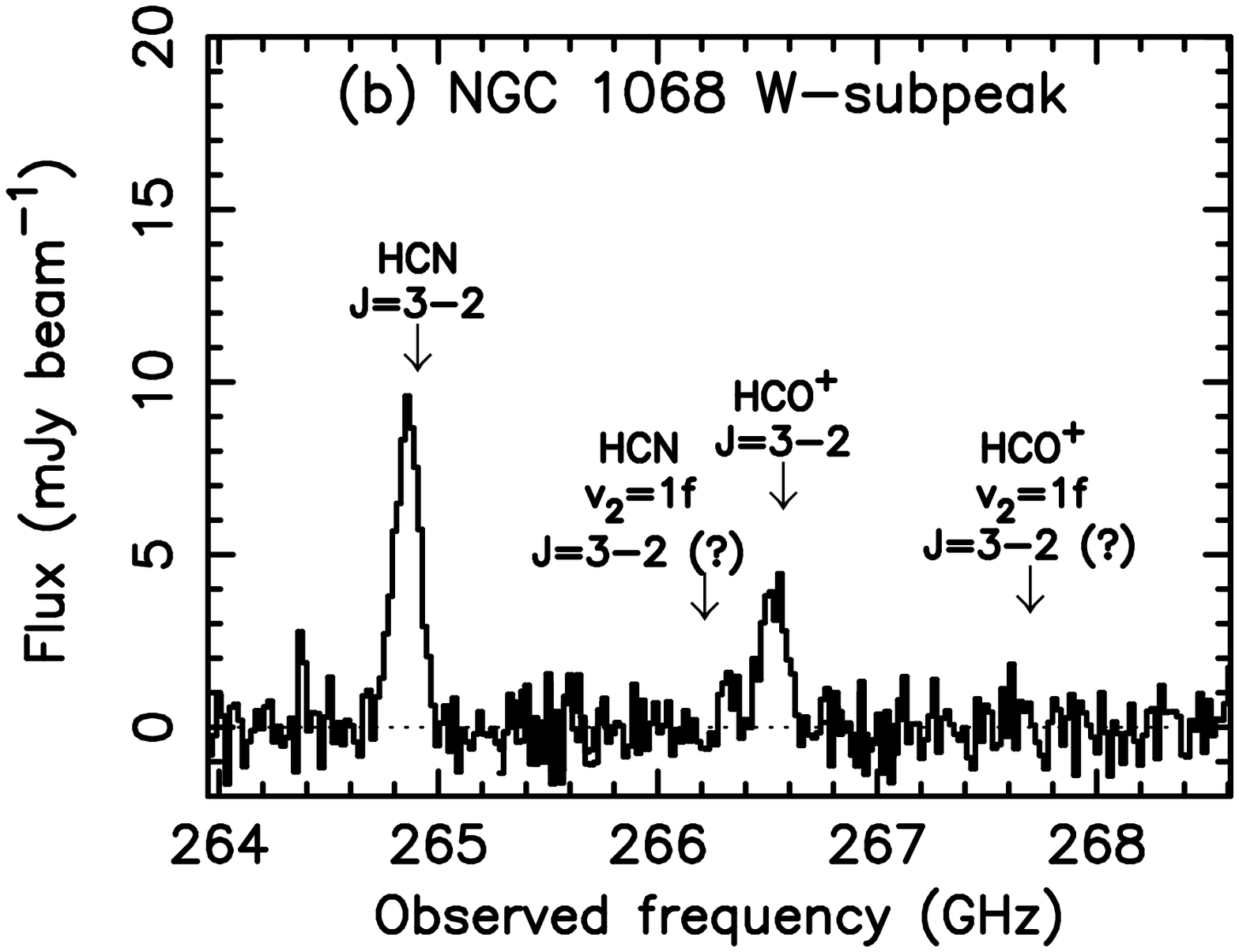} \\  
\includegraphics[angle=0,scale=.4]{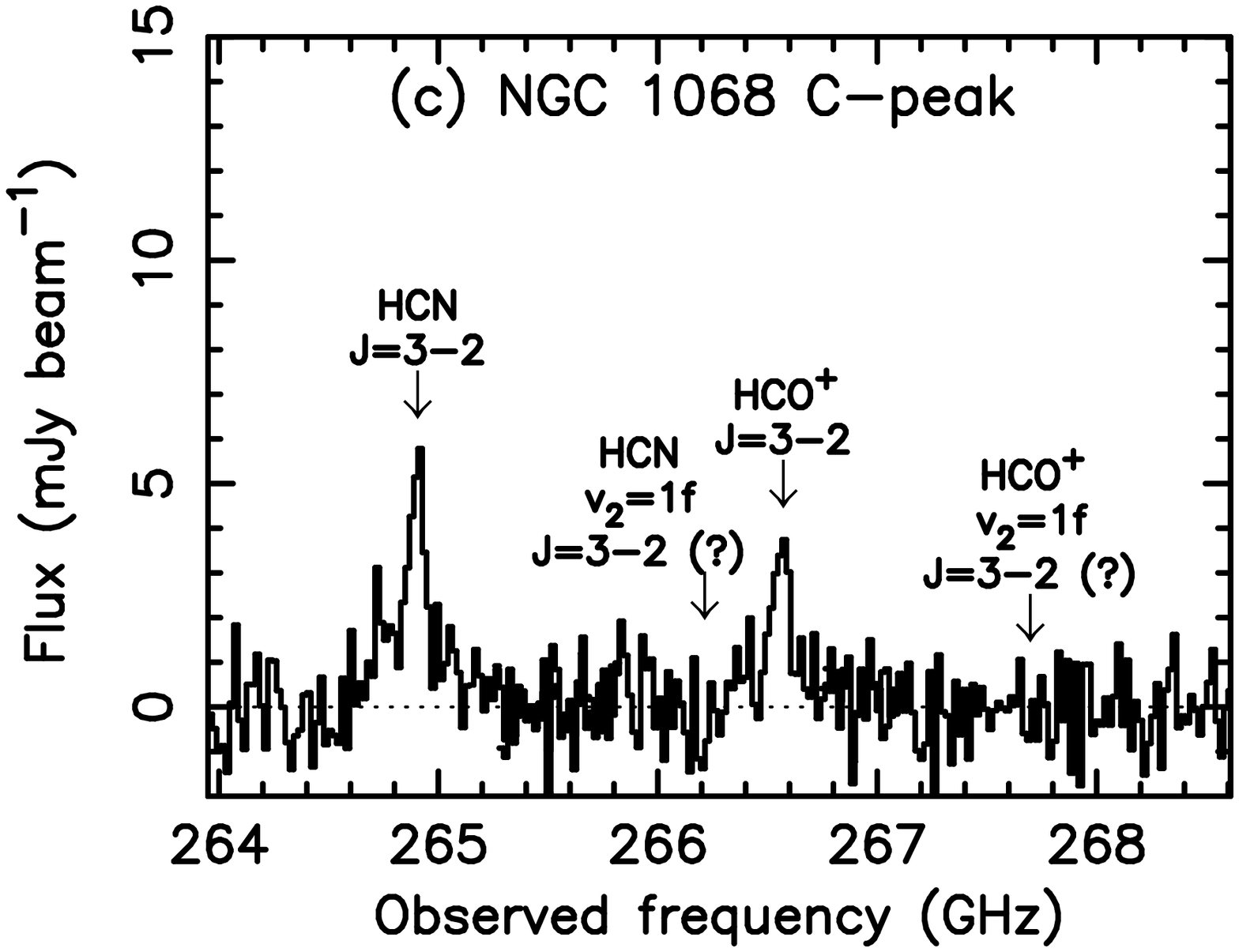} 
\includegraphics[angle=0,scale=.4]{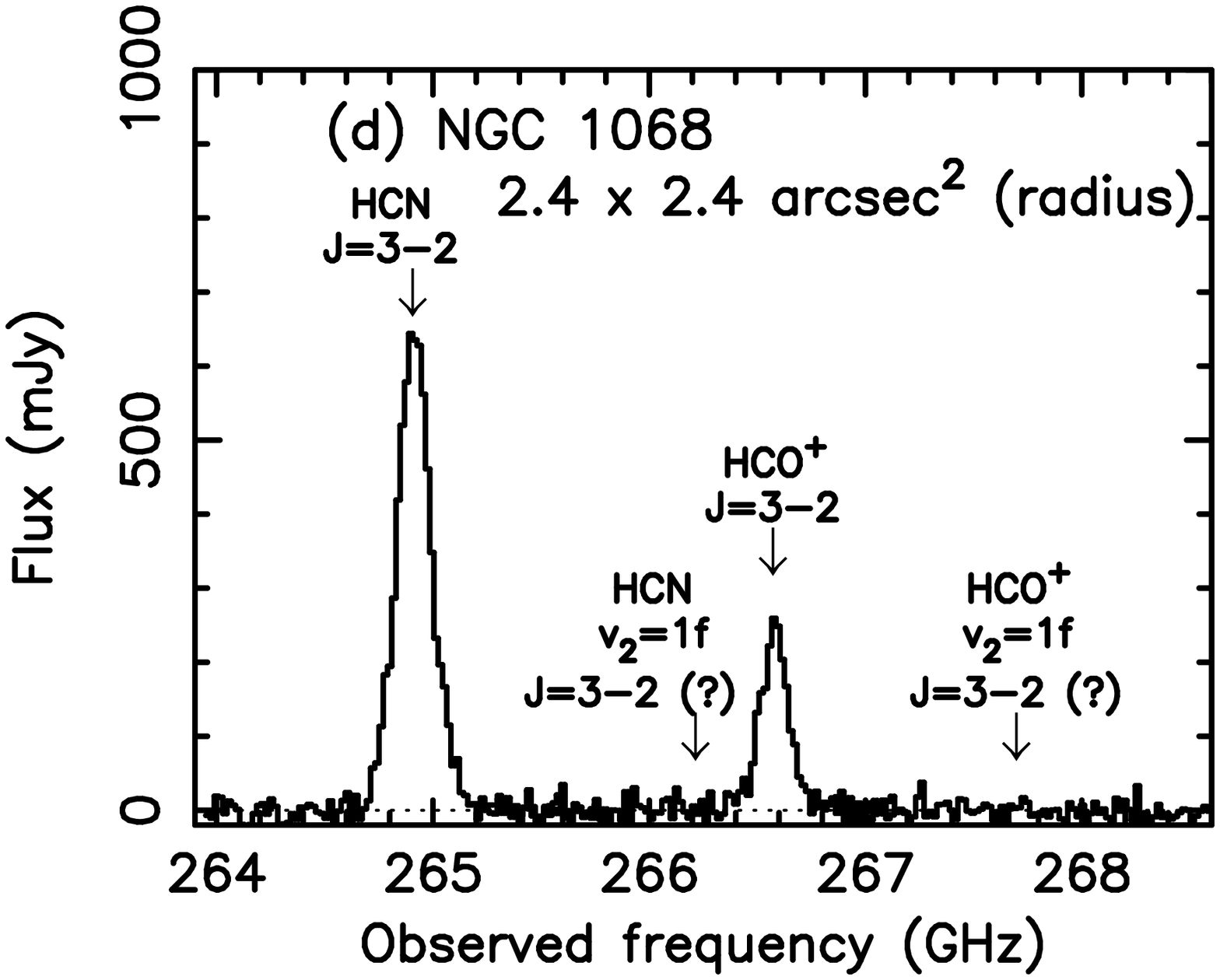} 
\end{center}
\vspace{-0.3cm}
\caption{Spectra, within the beam size, at interesting regions: 
(a) E-peak, (b) W-subpeak, and (c) C-peak.
An area-integrated spectrum with a 2$\farcs$4 radius circular region
centered at (02$^{h}$42$^{m}$40.69$^{s}$,$-$00$^{\circ}$00$'$48.50$''$)J2000 is also shown in (d).}
\end{figure}

\begin{figure}
\begin{center}
\includegraphics[angle=0,scale=.5]{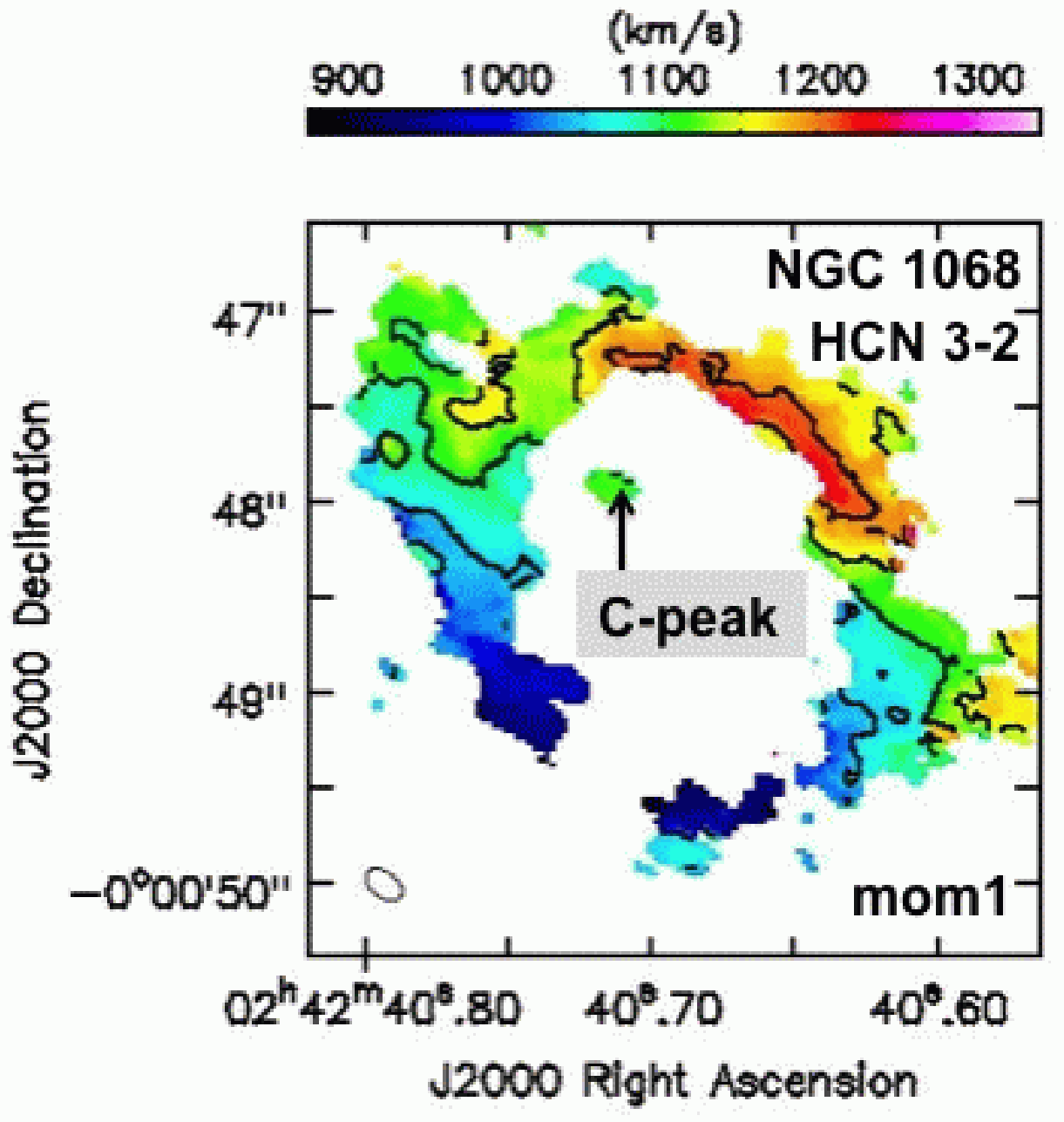} 
\includegraphics[angle=0,scale=.5]{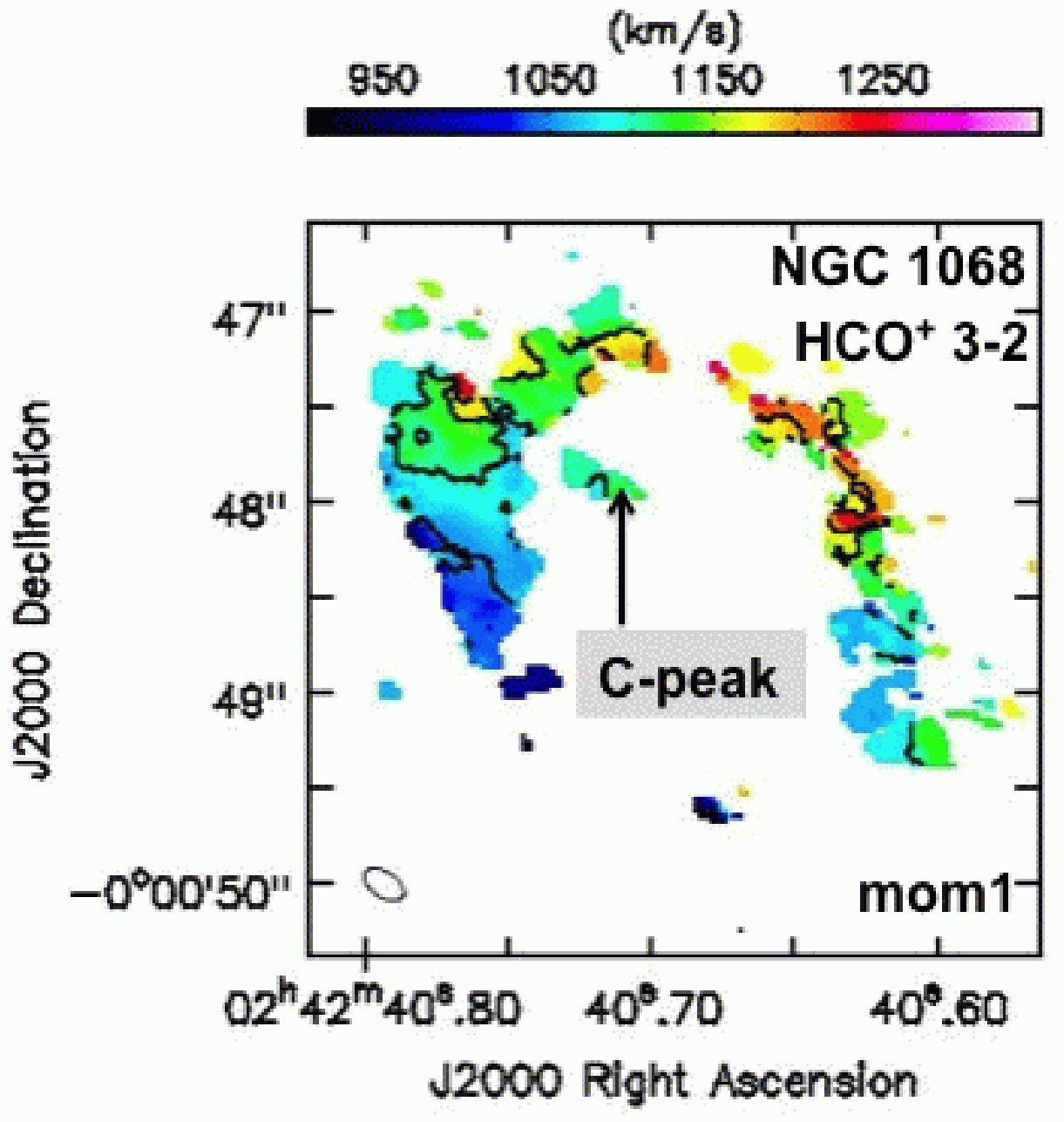} 
\vspace*{-0.3cm}
\end{center}
\caption{Intensity-weighted mean velocity (moment 1) maps of the HCN J=3--2 and
HCO$^{+}$ J=3--2 emission lines. 
The contours represent 1050, 1100, 1150, and 1200 km s$^{-1}$.} 
\end{figure}

\end{document}